\documentclass[letterpaper]{article}

\usepackage[T1]{fontenc}
\usepackage{geometry}
\usepackage{setspace}
\usepackage{graphicx}

\usepackage[articletitle=true,doi=true]{achemso}
\usepackage{etoolbox}
\usepackage{booktabs}
\usepackage{xcolor}
\usepackage{tikz}

\usepackage{graphicx}
\usepackage{float}
\newfloat{scheme}{htbp}{los}
\floatname{scheme}{Scheme}
\floatname{chart}{Chart}
\newfloat{graph}{htbp}{loh}

\usepackage{chemformula} 
\usepackage[version = 4]{mhchem} 

\usepackage{authblk}

\usepackage[
colorlinks=true,
citecolor=blue,
linkcolor=red,
urlcolor=blue,
bookmarks=false
]{hyperref}

\newcommand{\doi}[1]{\href{https://doi.org/#1}{\nolinkurl{#1}}}

\author[1]{Sumit Srivastav*}
\author[2]{Per Eng-Johnsson}
\author[3]{Raimund Feifel}
\author[1]{Alicja Domaracka}
\author[1]{Patrick Rousseau}
\author[4]{Dariusz G. Piekarski*}
\author[1]{Sylvain Maclot*}

\affil[1]{Université Caen Normandie, ENSICAEN, CNRS, CEA, Normandie Univ, CIMAP UMR6252, F-14000 Caen, France}
\affil[2]{Department of Physics, Lund University, P.O. Box 118, 22100, Lund, Sweden}
\affil[3]{Department of Physics, University of Gothenburg, Origovägen 6 B, 41296 Gothenburg, Sweden}
\affil[4]{Institute of Physical Chemistry, Polish Academy of Sciences, Kasprzaka 44/52, 01-224, Warsaw, Poland}
\title{Ionization-Driven Dehydrogenation and Molecular Growth in Gas-Phase Adamantane Clusters}
\date{*Email:~sumit.srivastav@unicaen.fr, dpiekarski@ichf.edu.pl, sylvain.maclot@ganil.fr}

\begin{document}

\maketitle

\begin{abstract}	
We investigate fragmentation and molecular growth induced by keV ions in initially neutral gas-phase adamantane clusters, the simplest diamondoid molecule (Ad, C$_{10}$H$_{16}$). Using time-of-flight mass spectrometry in combination with quantum-chemical calculations, we find that intact cluster cations undergo extensive dehydrogenation. We also observe a distinct distribution of singly charged products with masses predominantly between those of the adamantane monomer and dimer cations. We attribute these species to ion-induced growth products and interpret their formation considering the smallest cluster cation, [Ad]$_2^+$, using quantum-chemical calculations. Our results demonstrate that ionization strongly reorganizes the dimer geometry, facilitating H$_2$ elimination and intermolecular covalent C--C bond formation. Weakly bound adamantane clusters are thereby transformed into covalently bound carbonaceous species with a strong preference for specific molecular structures. These findings provide new insight into ionization-driven molecular growth in saturated hydrocarbon systems contributing to the understanding of molecular complexification, with implications for chemical physics and astrochemistry.
\end{abstract}

\section*{TOC Graphic}

\begin{center}
	\includegraphics[width=3.25in,height=1.75in]{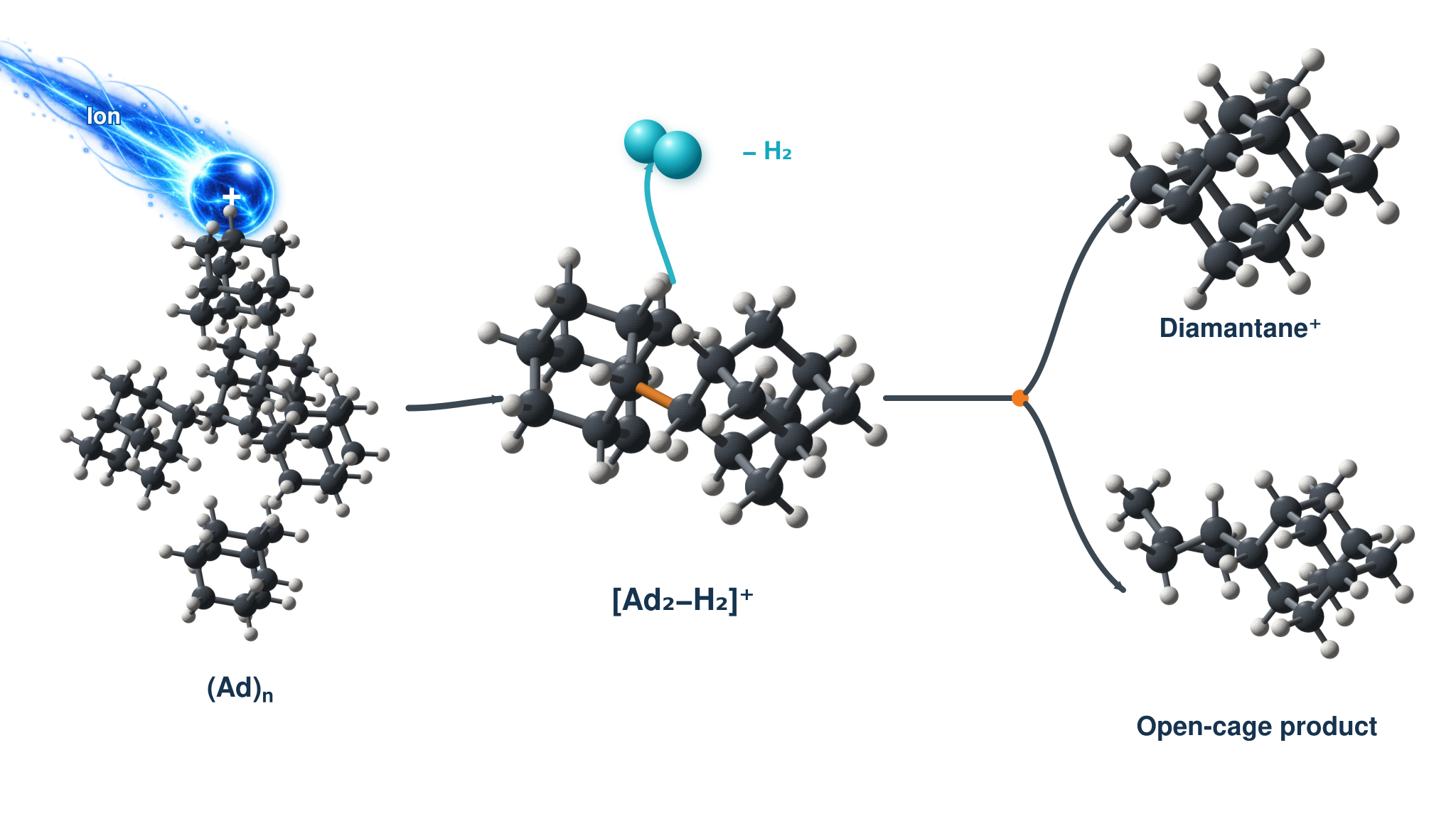}
\end{center}

\begin{figure*}[!ht]
	\centering
	\includegraphics[width=0.95\linewidth]{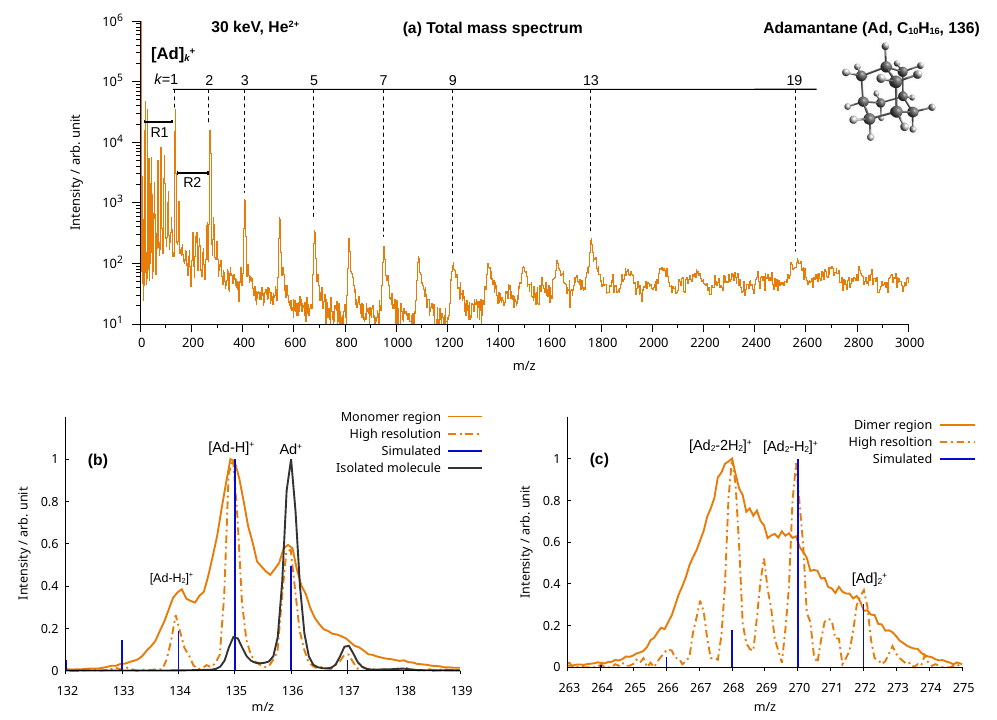}
	\caption{\label{fig:tot_del_simul}(a) Total mass spectrum of cationic products formed in collisions of neutral adamantane clusters with 30 keV He$^{2+}$. (b) and (c) Magnified views of the monomer ($k=1$) and dimer ($k=2$) mass regions, respectively. Solid orange curves show the corresponding experimental mass spectra, while dash-dotted orange curves show the high-resolution experimental spectra. Blue sticks represent the mass distributions obtained from molecular dynamics simulations. The black curve in panel (b) shows the mass spectrum of isolated adamantane measured under the same projectile ion. The ball-and-stick structure shown in panel (a) represents adamantane (Ad, C$_{10}$H$_{16}$, 136).}
\end{figure*}

Diamondoids constitute a distinctive class of saturated carbocyclic hydrocarbons because of their exceptional mechanical rigidity, thermal stability, and high symmetry.\cite{hartmut08} Adamantane (Ad, C$_{10}$H$_{16}$; inset of Figure~\ref{fig:tot_del_simul}(a)) is the smallest member of the diamondoid family. It consists entirely of hydrogen-terminated \textit{sp}$^3$-hybridized carbon atoms arranged in a three-dimensional diamond lattice structure. Diamondoids have attracted considerable interest owing to their unique structural and electronic properties.\cite{yang07,mochalin12}

The discovery of fullerenes in the interstellar medium (ISM)\cite{cami10,sellgren10,campbell15} motivated extensive searches for other stable carbonaceous species, including diamondoids and polycyclic aromatic hydrocarbons (PAHs).\cite{snow98,crandall20} More recently, PAHs have been unambiguously identified in the TMC-1 molecular cloud.\cite{mcGuire21} Diamondoids are also recognized as abundant constituents of presolar grains, and nanodiamond particles have been extracted from primitive meteorites.\cite{lewis87,jones04} Although isolated diamondoids have not yet been conclusively identified in the ISM, their remarkable structural and thermal stability suggests that they could survive in harsh astrophysical environments.\cite{henning98} In addition, the infrared emission bands at 3.43 and 3.53~$\mu$m observed toward several circumstellar disks have been attributed to nanodiamond particles,\cite{guillois99,habart04} with subsequent studies suggesting that even relatively small nanodiamonds containing up to about 130 carbon atoms could account for these features.\cite{pirali07}

Gas-phase investigations have allowed to obtain the ionization potentials and infrared/electronic spectra of diamondoids.\cite{lenzke07,oomens06,steglich11} These studies further suggested that the instability of ionized diamondoids may explain why isolated diamondoids have not yet been identified in the ISM. Consequently, the fragmentation dynamics of isolated adamantane cations have been investigated using photoionization, electron ionization, ultrafast XUV excitation, infrared spectroscopy, and molecular dynamics simulations.\cite{candian18,bouwman18,maclot20,ganguly22,roy23,ganguly23} Together, these studies demonstrated that ionization can destabilize the nominally rigid diamondoid cage, induce hydrogen loss and cage opening, and produce smaller hydrocarbon fragments.

Ionizing radiation is expected to play an important role in molecular growth throughout astrophysical and planetary environments by driving both bottom-up assembly and top-down chemical evolution of carbonaceous matter.\cite{dunk12,berne12,micelotta12,delaunay15,jäger09,shuman15,cable12,srivastav25} Accordingly, ion-induced intracluster reactions have been extensively investigated in fullerenes,\cite{zettergren10,zettergren13,seitz13,heden05} PAHs,\cite{holm10,seitz11,delaunay15,nair24} and mixed fullerene--PAH clusters,\cite{alicja18} where collisions produce a wide range of molecular growth products through intracluster bond formation (ICBF). While these systems are dominated by conjugated \textit{sp}$^2$ carbon frameworks, diamondoids represent fully saturated three-dimensional \textit{sp}$^3$ hydrocarbons that may exhibit fundamentally different reaction pathways. Despite their potential astrophysical relevance, experimental studies addressing molecular growth in diamondoid clusters remain scarce. Recent helium nanodroplet experiments have characterized adamantane cluster ions, their dehydrogenated derivatives, and mixed adamantane--water clusters,\cite{kappe22,goulart17,martin23,lorenz18} but ionization-driven intracluster growth in diamondoid clusters has remained essentially unexplored.

In the present work, we investigate ion-induced fragmentation and the formation of new molecular growth products in ionized adamantane clusters produced after collisions with keV-energy projectile ions. Using time-of-flight mass spectrometry combined with quantum-chemical calculations, we show that ionization of neutral adamantane clusters results in extensive dehydrogenation of the corresponding cluster cations. In addition, we observe a distinct distribution of singly charged higher-mass species, predominantly located between the adamantane monomer and dimer cations, demonstrating ion-induced intracluster reactivity. These species are assigned to covalently bound growth products formed through ICBF process. To elucidate these experimental observations, we combine molecular dynamics (MD) simulations with potential energy surface (PES) calculations at density functional theory (DFT) and semi-empirical levels of theory for singly charged dimer, [Ad]$_2^+$. We also highlight that the mechanisms governing intracluster reactivity in adamantane clusters are different from those previously reported for PAH and fullerene clusters.

Figure~\ref{fig:tot_del_simul}(a) shows the total mass spectrum obtained after collisions of 30~keV He$^{2+}$ ions with neutral adamantane clusters. Peak labeled [Ad]$_k^+$ ($1\leq k\leq19$) indicate the nominal mass positions associated with the adamantane monomer ($k=1$) and the series of $k$-mer cluster cations. This cluster distribution is produced predominantly by sequential evaporation of neutral adamantane molecules from larger clusters.
The intermolecular binding energies of clusters cations, [Ad]$_k^+$, are well below 1~eV and are tabulated in the \textcolor{blue}{Supporting Information}. The enhanced abundances of [Ad]$_{13}^{+}$ and [Ad]$_{19}^{+}$ correspond to the ``magic-number'' clusters associated with particularly stable packing arrangements previously observed experimentally and reproduced theoretically.\cite{goulart17,javier19} Products below the adamantane monomer mass (region R1) arise from fragmentation of strongly excited individual molecules within clusters following close or penetrating interaction with the projectile ion. In contrast, the intermediate species observed predominantly between the monomer and dimer cations (region R2) cannot be explained by simple evaporation of intact adamantane molecules and are assigned to intracluster reactions leading to the formation of new covalently bound hydrocarbon ions.

To further characterize the fragmentation and reactivity of produced adamantane cluster cations, zoomed-in views of the monomer and dimer regions of the total mass spectrum are shown in Figures~\ref{fig:tot_del_simul}(b) and \ref{fig:tot_del_simul}(c), respectively. 
The dotted curves represent the corresponding high-resolution spectra obtained using time-lag focusing technique (see experimental section). To identify the molecular ion position in the cluster fragmentation spectrum, we also displayed the spectrum of isolated gas-phase adamantane obtained under the same collision conditions, and the corresponding molecular ion peaks are shown by the black curves (Figure~\ref{fig:tot_del_simul}(b)). For isolated adamantane, Ad$^+$ is the dominant product, while dehydrogenated [Ad-H]$^+$ appears as a weaker secondary fragment. The peak at $m/z$  = 137 arises from the natural abundance of the $^{13}$C isotope. In contrast, the cluster fragmentation into the monomer region is dominated by [Ad-H]$^+$ at $m/z$  = 135, followed by Ad$^+$ and [Ad-H$_2$]$^+$ at $m/z$  = 136 and $m/z$  = 134, respectively. This change indicates that the cluster environment, and/or fragmentation of larger ionized aggregates strongly enhances dehydrogenation. The effect becomes even more pronounced in the dimer region (Figure~\ref{fig:tot_del_simul}(c)), however the mechanism related with H loss is different. Although multiple dehydrogenation is also observed for larger cluster cations, we focus here on the monomer and dimer regions to understand the underlying mechanism. For the dimer region, the dominant product is [Ad$_2$-2H$_2$]$^+$ at $m/z$  = 268, and losses of up to six hydrogen atoms or three hydrogen molecules are also observed.  In previous work, comparable adamantane clusters formed in helium droplets exhibited considerably less extensive dehydrogenation, with fragment intensities generally decreasing as the number of hydrogens loss increases.\cite{goulart17} The present experiment exhibits a different fragmentation pattern, reflecting the different excitation and relaxation regime following keV ion impact. The multiple dehydrogenation may proceed either through the loss of molecular H$_2$ or through sequential emission of H atoms. To understand dehydrogenation pathways and quantify their relative contributions, a comprehensive MD simulations and PES calculations were therefore performed.

The simulated mass spectra for the monomer and dimer regions are shown  as blue vertical lines in Figures~\ref{fig:tot_del_simul}(b) and \ref{fig:tot_del_simul}(c), respectively. The simulations reproduce the experimental monomer region well, whereas noticeable deviation remains in the dimer region, as described in the following discussion. The MD simulations were performed for the adamantane dimer, considering three distinct dimer cation structures at various internal excitation energies with focus on trajectories leading to dehydrogenated products (see Table~S2 for details). The interplay between atomic H and molecular H$_2$ loss was analyzed, and atomic H emission is found to be rare, occurring in only 9 of the 1200 trajectories. In contrast, 54.9\% of the trajectories release at least one H$_2$ molecule, and exactly one H$_2$ is emitted in 45.7\% of the full ensemble. Two distinct H$_2$-formation pathways were identified: an intramolecular pathway, in which both H atoms originate from the same adamantane residue, and an intermolecular pathway, in which the two H atoms originate from different adamantane residues. Interestingly, the intermolecular pathway accounts for 64.3\% of all H$_2$ formation events. The first H$_2$ emission is particularly selective, where among the trajectories releasing at least one H$_2$, 81.8\% of the first H$_2$-loss events proceed through the intermolecular pathway. In trajectories with several H$_2$ losses, the intramolecular contribution further starts increasing (see Tables~S2 and S3). The PES analysis of the adamantane dimer further examines the energetics associated with both H$_2$-formation pathways.

\begin{figure*}[t]
	\centering
	\includegraphics[width=0.9\textwidth]{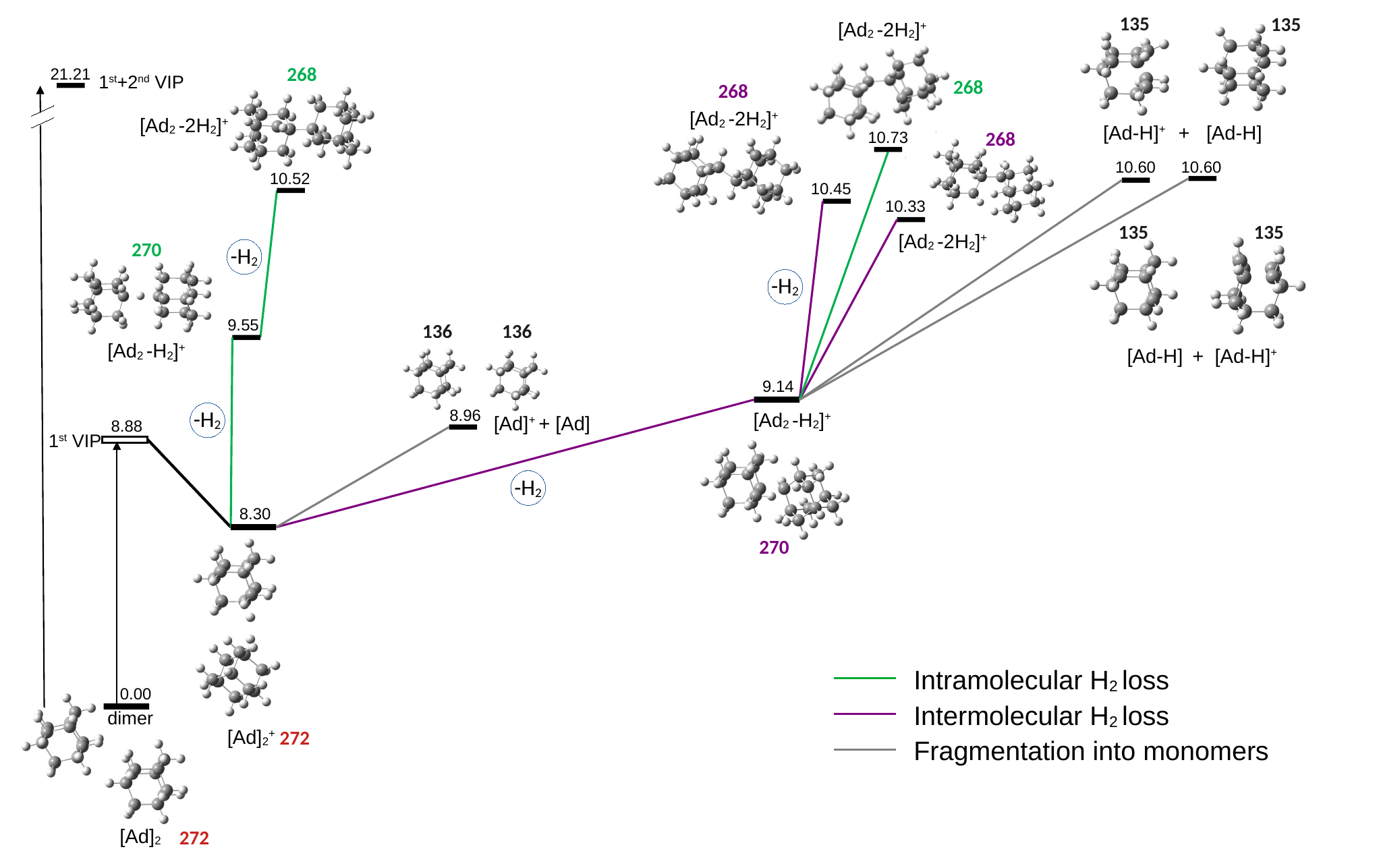}
	\caption{\label{fig:PES_H2}Potential energy surface for H$_2$ elimination and subsequent fragmentation pathways of the singly charged adamantane dimer. Relative energies were calculated at the B3LYP-D3(BJ)/6-311++G(d,p) level with respect to the neutral dimer and are given in eV.}
\end{figure*}

The dehydrogenation pathways of the adamantane dimer cation are summarized in Figure~\ref{fig:PES_H2}. Following vertical ionization at 8.88~eV, [Ad]$_2^+$ relaxes to its adiabatic minimum at 8.30~eV, corresponding to the most stable singly charged dimer identified in this work (see Figures S2 and S3 for all neutral and charged isomers). Both the intramolecular and intermolecular pathways for dehydrogenation were explored. Ionization substantially reorganizes the bonding region between the two weakly bound adamantane residues. In particular, the neighboring C--H bonds lengthen by approximately 0.04 \AA, while the H$\cdots$H separation decreases from 2.42~\AA~to 1.44~\AA. This structural reorganization preorganizes the adiabatic cation for H$_2$ elimination by activating the neighboring C--H bonds. Intermolecular H$_2$ loss subsequently forms a covalently bound [Ad$_2$-H$_2$]$^+$ structure at 9.14~eV, thus enables direct intermolecular C--C bond formation following the first H$_2$ elimination. Subsequent dehydrogenation favors products derived from the intermolecular pathway; the covalently bound structures formed after a second H$_2$ loss lie at 10.33/10.45~eV, compared with 10.52/10.73~eV for the corresponding intramolecular products. The quantum-chemical calculations therefore suggest a mechanistic sequence in which ionization first activates two intermolecular C--H bonds and favors H$_2$ formation. Subsequent dehydrogenation and structural rearrangement open additional intramolecular pathways. Interestingly such dehydrogenation channels have not been reported for PAH cluster cations produced from the same cluster source under collisions with various projectile ions.\cite{delaunay15,alicja18,delaunay18,rousseau12}

The weakly bound dimer cation may alternatively dissociate into Ad$^+$+Ad at 8.96~eV. Moreover, after the loss of one H$_2$ molecule at 9.14~eV, fragmentation into [Ad-H]$^+$ + [Ad-H] occurs at 10.60~eV, with nearly isoenergetic products corresponding to charge localization on either adamantane unit. This pathway provides a direct connection between H$_2$ elimination in the dimer and the intense [Ad-H]$^+$ experimental signal observed in the monomer region. The calculations further show that the initial [Ad$_2$-H$_2$]$^+$ structure strongly influences the dynamics. Trajectories initiated from this covalently bound minimum at 9.14~eV preferentially retain the dimer skeleton and undergo further H$_2$ elimination, producing strong simulated contribution to peaks at $m/z=268$, 266, and 264. In contrast, trajectories started from the noncovalent adiabatic minimum more frequently dissociate after the loss of one or two H$_2$ molecules, contributing predominantly to the $m/z=135$ peak. The significant observation of $m/z$  = 268 in both the experiment and the covalently bound trajectory ensemble is therefore consistent with stabilization through H$_2$-assisted intermolecular C--C bond formation. 

\begin{figure}[!b]
	\centering
	\includegraphics[width=0.7\linewidth]{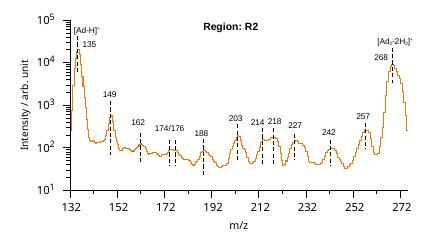}
	\caption{\label{fig:growth}Magnified view of region R2 of the total mass spectrum shown in Figure~\ref{fig:tot_del_simul}(a), showing the singly charged products observed between the adamantane monomer and dimer cation regions. The labeled peaks indicate the nominal $m/z$ values of the observed features.} 
\end{figure}

The quantum-chemical calculations were limited to the dimer, whereas the experimentally observed multiply dehydrogenated products in the dimer mass region may also originate from larger clusters. Such interpretation is supported by the simulated monomer spectrum, where multiple H$_2$ losses arise from fragmentation of dehydrogenated dimer cationic species. Consequently, the simulations reproduce the experimental monomer spectrum well but underestimate the intensities of the multiply dehydrogenated products in the dimer region (see Figures~\ref{fig:tot_del_simul}(b) and \ref{fig:tot_del_simul}(c)). Moreover, our ADMP simulations at semi-emiprical level underestimates H atom loss comparing with DFT-based Born-Oppenheimer Molecular Dynamics (BOMD) results for smaller hydrocarbons.\cite{kumari25} Despite these limitations, the simulations capture the principal experimental trends and provide a consistent mechanistic picture of ionization-driven dehydrogenation and intracluster bond formation in adamantane clusters. 

Figure~\ref{fig:growth} shows an enlarged view of region R2, located between the monomer and dimer cations in the total mass spectrum shown in Figure~\ref{fig:tot_del_simul}(a). The intermediate species observed in this region indicate that fragmentation and covalent molecular growth compete within the excited cluster. Their survival on the microsecond timescale of the experiment further suggests that at least part of this distribution corresponds to (meta)stable covalently bound ions rather than short-lived collision complexes.

Previous studies of PAH and fullerene clusters have shown that ion-induced intracluster reactivity is initiated by nuclear stopping during ion collisions, causing the prompt knockout of one or more atoms from an individual molecule.\cite{delaunay15,zettergren13} The resulting reactive fragments subsequently undergo bond formation with neighboring molecules within the cluster, leading to molecular growth. Accordingly, the growth-product yields have exhibited a strong dependence on projectile mass (see Figure~2 of Ref.~\cite{delaunay15}), with little or no growth observed for light projectiles such as H$^+$ and He$^+$. In contrast, we did not observe such a pronounced projectile-mass dependence for adamantane clusters. Rich growth-product distributions were measured for both the light He$^{2+}$ projectile and the heavier O$^{3+}$ and Kr$^{11+}$ ions (see Figure~S1). Nevertheless, the relative abundances of the growth products varied with projectile ions, reflecting differences in the internal excitation of the cluster cations. These observations indicate that intracluster reactivity in adamantane clusters proceeds through a mechanism distinct from the knockout-driven processes established for PAH and fullerene clusters. Unlike these highly conjugated systems, which can store substantial internal excitation energy without extensive fragmentation, adamantane breaks efficiently following ionization by external perturbations.\cite{candian18,bouwman18,maclot20} Consequently, ionization-driven molecular fragmentation appears to be the dominant pathway initiating reactivity within adamantane clusters.

\begin{figure*}[!t]
	\centering
	\includegraphics[width=0.8\linewidth]{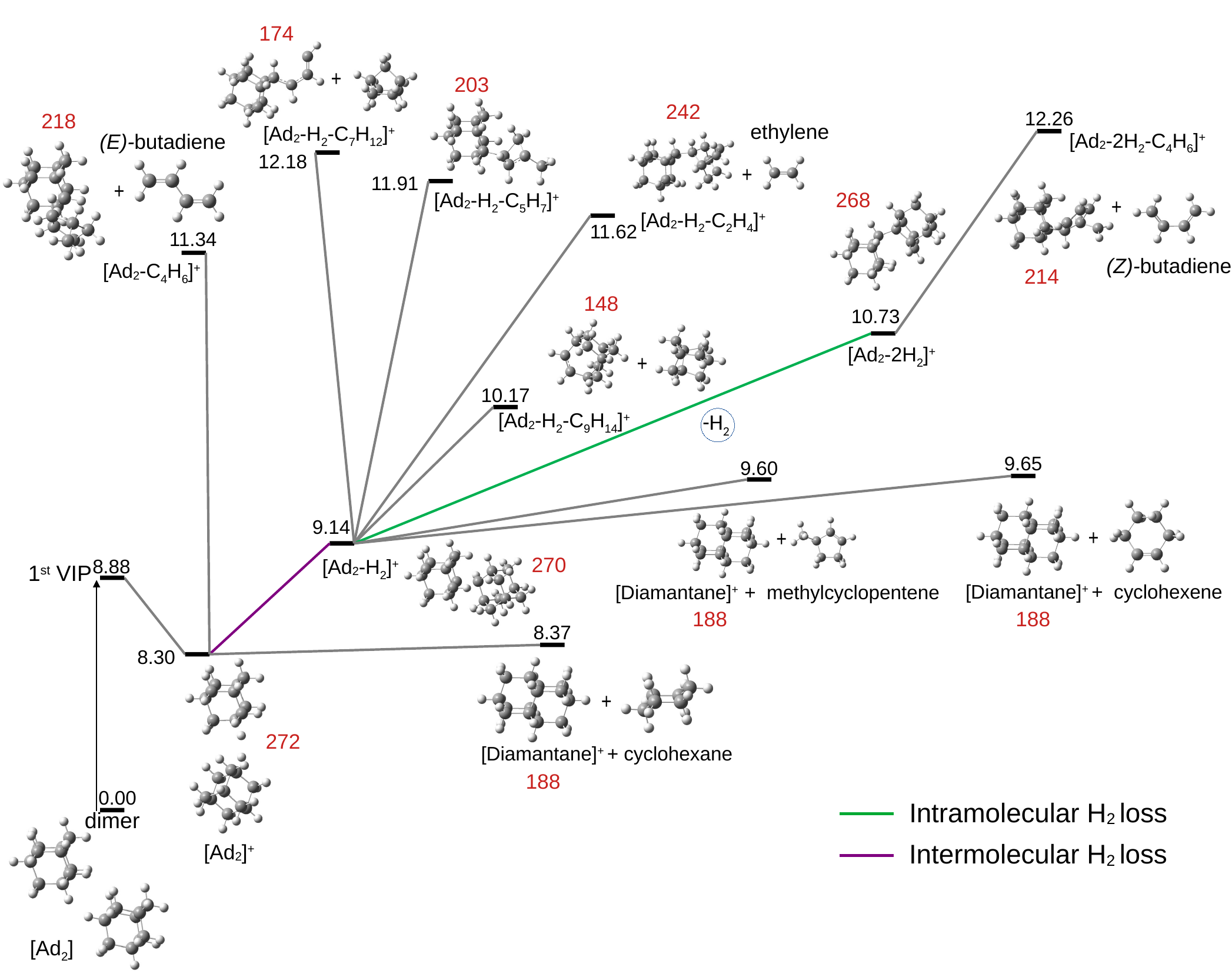}
	\caption{\label{fig:PES1_growth}Potential energy surface for selected experimentally observed products, considering H$_2$ elimination as the initial step of the dynamics followed by neutral emission. The $m/z=188$ (8.37 eV) and $m/z=218$ (11.34 eV) products proceed through direct neutral emission without prior H$_2$ elimination. Relative energies were calculated at the B3LYP-D3(BJ)/6-311++G(d,p) level with respect to the neutral adamantane dimer and relative zero point corrected electronic energies are given in eV.} 
\end{figure*}

Although contributions from larger clusters cannot be excluded, many of the observed growth products can be rationalized by direct fragmentation of the adamantane dimer cation following sequential neutral emission. Figure~\ref{fig:PES1_growth} summarizes the calculated reaction pathways, considering H$_2$ elimination as the initial step of the dynamics. The products at $m/z=214$ and 242 are assigned to (\textit{Z})-butadiene emission from [Ad$_2$-2H$_2$]$^+$ at 12.26~eV and ethylene emission from [Ad$_2$-H$_2$]$^+$ at 11.62~eV, respectively. Similarly, the peaks at $m/z=148$, 174, and 203 can be attributed to the emission of larger neutral hydrocarbon fragments at 10.17, 12.18, and 11.91~eV, respectively. The $m/z=188$ (8.37 eV) and $m/z=218$ (11.34 eV) products, however, are obtained by direct neutral emission of cyclohexane and (\textit{E})-butadiene, respectively, from the adiabatic minimum of [Ad]$_2^+$ at 8.30~eV, without prior H$_2$ elimination. Interestingly, the peak at $m/z=188$ coincides with the molecular mass of diamantane, the second member of the diamondoid family. The formation of the diamantane cation seems to be thermodynamically accessible through the elimination of neutral cyclohexane at 8.37~eV. Following H$_2$ loss, the same product can also be formed through the emission of either cyclohexene (9.65~eV) or a five-membered ring fragment (9.60~eV). It suggests that such ionization-driven chemistry can contribute to growth toward higher-order diamondoids
in carbonaceous matter under astrophysical conditions.

\begin{figure*}[!t]
	\centering
	\includegraphics[width=0.8\linewidth]{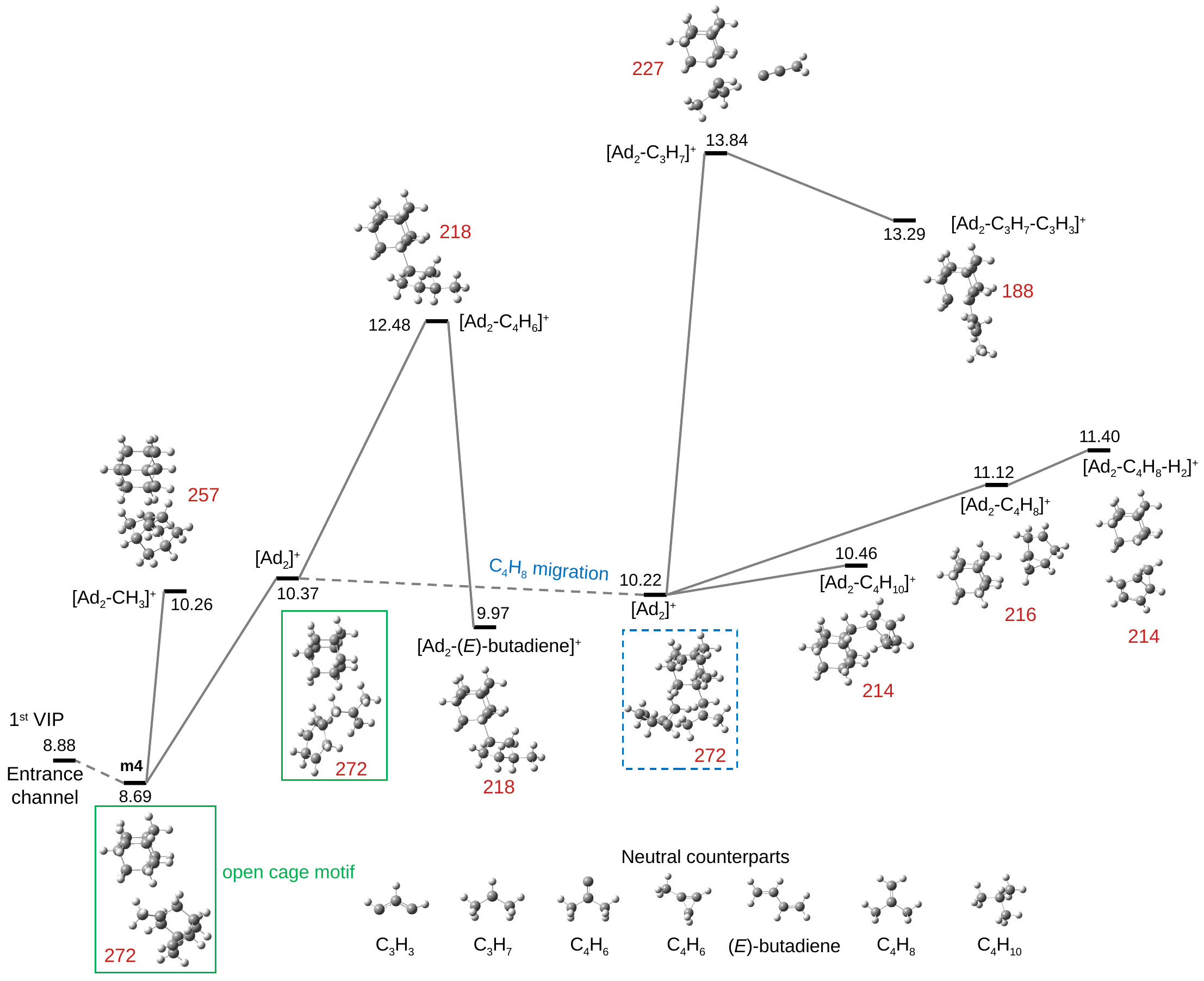}
	\caption{\label{fig:PES2_growth}Potential energy surface for selected pathways involving cage opening and hydrocarbon-chain migration between the two adamantane residues. Relative energies were calculated at the B3LYP-D3(BJ)/6-311++G(d,p) level with respect to the neutral dimer. Relative zero point corrected electronic energies are given in eV. The structures of released neutrals are given at the bottom. } 
\end{figure*}

Molecular adamantane cations are known to undergo efficient fragmentation following cage opening.\cite{candian18} Analogously, we explored a complementary mechanism in which one adamantane cage of dimer cation first opens before undergoing further rearrangement, giving rise to growth products (see Figure~\ref{fig:PES2_growth}). Compared with the compact covalently bound dimer pathway discussed above, this mechanism lowers the energy required to reach some products, particularly the $m/z=214$ channel, which becomes accessible at 10.46 and 11.40~eV. The reaction is initiated from the adiabatic dimer cation through formation of an open-cage intermediate via consecutive transition states involving cage opening, CH$_3$ edge formation, and H-migration. A detailed analysis of the cage-opening, CH$_3$ edge formation, and H-migration, is provided in Figure~S6 of the \textcolor{blue}{Supporting Information}. The resulting open-cage structures contain both methylene-like and olefinic motifs, providing a structural link between the initially saturated diamondoid framework and mixed \textit{sp$^3$/sp$^2$} polycyclic carbonaceous ions. Such processes may contribute to the structural evolution of carbonaceous matter exposed to energetic radiation in astrophysical environments.\cite{hrodmarsson25,pelaez18,herrero22} We further examined migration of a C$_4$H$_8$ fragment between the two residues, leading to more stable minimum at 10.22~eV, which by direct bond breaking can connect to several additional reaction pathways. Within this mechanism, alternative $m/z=188$ isomers, distinct from diamantane, were identified at 13.29~eV, substantially higher in energy than diamantane.

The PES analysis above focused on reaction pathways involving structures lying within approximately 5~eV above the vertical ionization potential of [Ad$_2$]$^+$ (8.88~eV). This energy window is consistent with the peak of internal-energy distribution deposited following keV-ion collisions, with the distribution extending up to approximately 20~eV.\cite{ewa21,maclot16} Some experimentally observed peaks, including those at $m/z=162$ and 174 (although weak), as well as alternative fragmentation pathways leading to $m/z=188$ and 227, are associated with considerably higher-energy regions of the PES (see extended PES, Figures~S8 and S9, in the \textcolor{blue}{Supporting Information}). These products may therefore originate from larger clusters and/or proceed through reaction pathways that are beyond the scope of the present dimer model.

In summary, we demonstrated that keV-ion collisions with neutral adamantane clusters induce ionization and excitation followed by extensive dehydrogenation and intracluster covalently bound molecular growth, yielding long-lived ionic products that survive for at least microsecond timescale. The dehydrogenation resulted in [Ad-H]$^+$ becoming the dominant product in the adamantane monomer region, while [Ad$_2$-2H$_2$]$^+$ was the most intense species in the corresponding dimer region. Molecular dynamics simulations reproduced the pronounced dehydrogenation and showed that molecular H$_2$ emission was considerably more probable than atomic H loss. The first H$_2$ molecule is formed predominantly through an intermolecular mechanism involving one hydrogen atom from each adamantane residue. Quantum-chemical calculations further established that ionization activates the neighboring C--H bonds, enabling intermolecular H$_2$ elimination followed by C--C bond formation between the two adamantane residues in the dimer cations [Ad]$_2^+$, and stabilizing covalently bound dehydrogenated dimers. Subsequent neutral fragment emission from these covalently bound intermediates accounted for many of the experimentally observed growth products. A complementary cage opening pathway provided additional energetically accessible routes to larger carbonaceous growth products, including mixed \textit{sp}$^3$/\textit{sp}$^2$ polycyclic structures. Unlike the knockout-driven molecular growth established for PAH and fullerene clusters, the present results indicated that intracluster reactivity in adamantane clusters is primarily initiated by ionization-driven fragmentation. 
Ionized adamantane clusters therefore provide a model system for understanding how weakly bound aggregates of hydrogen-rich symmetric complex carbon cage  can undergo dehydrogenation, covalent bond formation, and structural reorganization following energetic processing. More broadly, we show that energetic processing of saturated carbonaceous material can drive a transition from \textit{sp}$^3$-rich frameworks toward increasingly unsaturated and structurally diverse mixed \textit{sp}$^3$/\textit{sp}$^2$ carbon networks. The energetically accessible formation of diamantane from the adamantane dimer further suggests that such ionization-driven chemistry could contribute to bottom-up growth toward higher-order diamondoids, providing a possible molecular route toward greater structural complexity in carbonaceous matter under astrophysical conditions.
\subsection{EXPERIMENTAL METHODS}
The experiments were performed at the low-energy ion beam infrastructure ARIBE of the GANIL facility in Caen, France.\cite{bernigaud2008aribe} A detailed description of
the setup and the experimental method is given elsewhere.\cite{bergen1999CLUSTERS} Bunched ion beams were produced in an electron cyclotron resonance ion source, providing ion pulses of 0.5~$\mu$s in length and with a repetition frequency of 1.7~kHz. The ions used were 30~keV He$^{2+}$, 48~keV O$^{3+}$ and 165~keV Kr$^{11+}$. In a crossed-beam configuration, the ion beam interacts with an effusive beam of neutral clusters with a wide size distribution centered around 20-30 molecules. The clusters are formed in a gas aggregation source, where a vapor mixture of adamantane molecules (produced by evaporation of powder) and helium carrier gas is transported through a gas injection line into a liquid-nitrogen-cooled nozzle. Adamantane (Ad) has the molecular formula C$_{10}$H$_{16}$ and a molecular mass of 136~u. The sample was a commercial powder from Sigma-Aldrich ($\ge$ 99$\%$ purity) heated around 333~K with the gas injection line gradually heated to 373~K. After ion interaction, the reaction products are mass-over-charge analyzed in a linear time-of-flight spectrometer.\cite{chandezon1994WilerMcLaren} In order to resolve the broad mass distributions, the time-lag focusing conditions were implemented in selected mass regions by varying the delay of extraction with respect to the pulsed ion-beam.\cite{wiley55} Using this method, we achieved unit-mass resolution for the intact cationic adamantane clusters. Concerning the cases of isolated molecules, the experiments have been performed on the same spectrometer. The only difference is that the effusive jet of neutral molecules were produced with a heated gas injection line terminated by a needle. 
\subsection{COMPUTATIONAL DETAILS}
Potential energy surfaces were calculated with density functional theory using B3LYP\cite{becke93,lee88,becke96} and the 6-311++G(d,p) basis set.\cite{mclean80,krishnan80} Dispersion interactions were included using the D3 correction with Becke-Johnson damping,\cite{grimme10,grimme11} and the calculations were performed with Gaussian~16.\cite{g16} The energies reported in the potential energy surfaces are relative to the neutral adamantane dimer unless stated otherwise.

Production molecular dynamics simulations were carried out in Gaussian~16 using atom-centered density-matrix propagation (ADMP)\cite{iyengar01,schlegel01,schlegel02} with the PM7R8 semiempirical Hamiltonian and its hydrogen bonding correction.\cite{throssel17} Internal excitation energies of 15, 20, 25, and 50 eV were randomly distributed among the vibrational degrees of freedom. Three types of initial structures were considered: (i) the most stable neutral dimer geometry after removal of one outer electron, representing a Franck-Condon ionization; (ii) the optimized adiabatic dimer cation at 8.30 eV (see minimum in Fig.~\ref{fig:PES_H2}); and (iii) the covalently bound [Ad$_2$-H$_2$]$^+$ structure at 9.14 eV formed after intermolecular H$_2$ loss. In total, 1200 trajectories were propagated for 1000 fs with a 0.25 fs time step and an electronic fictitious mass of 0.15 amu. All simulations were performed for the cation in its lowest spin ground electronic state.

For trajectory analysis, fragments were assigned from the evolving bonding pattern with 2.5 \AA~for bond-breaking cutoff. H$_2$ formation was classified as intramolecular when both hydrogen atoms originated from the same adamantane residue and intermolecular when the two atoms originated from different residues. The simulated mass spectra were obtained by collecting the final charged fragments from all production runs for all excitation energies. 

\subsection{AUTHOR INFORMATION}
\subsection{Corresponding Authors}
\textbf{Sumit Srivastav~--~}Université Caen Normandie, ENSICAEN, CNRS, CEA, Normandie Univ, CIMAP UMR6252, F-14000 Caen, France; \href{https://orcid.org/my-orcid?orcid=0000-0001-9208-8051}{orcid.org/0000-0001-9208-8051};
Email: \href{mailto:sumit.srivastav@unicaen.fr}{sumit.srivastav@unicaen.fr}\\
\textbf{Dariusz G. Piekarski~--~}Institute of Physical Chemistry, Polish Academy of Sciences, Kasprzaka 44/52, 01-224, Warsaw, Poland; \href{https://orcid.org/0000-0003-1220-1902}{orcid.org/0000-0003-1220-1902};
Email: \href{mailto:dpiekarski@ichf.edu.pl}{dpiekarski@ichf.edu.pl}\\
\textbf{Sylvain Maclot~--~}Université Caen Normandie, ENSICAEN, CNRS, CEA, Normandie Univ, CIMAP UMR6252, F-14000 Caen, France; \href{https://orcid.org/0000-0001-5587-7182}{orcid.org/0000-0001-5587-7182};
Email: \href{mailto:sylvain.maclot@ganil.fr}{sylvain.maclot@ganil.fr}

\subsection{Authors}
\textbf{Per Eng-Johnsson~--~}Department of Physics, Lund University, P.O. Box 118, 22100, Lund, Sweden\\
\textbf{Raimund Feifel~--~}Department of Physics, University of Gothenburg, Origovägen 6 B, 41296 Gothenburg, Sweden\\
\textbf{Alicja Domaracka~--~}Université Caen Normandie, ENSICAEN, CNRS, CEA, Normandie Univ, CIMAP UMR6252, F-14000 Caen, France; \href{http://orcid.org/0000-0003-1585-2436}{orcid.org/0000-0003-1585-2436}\\
\textbf{Patrick Rousseau~--~}Université Caen Normandie, ENSICAEN, CNRS, CEA, Normandie Univ, CIMAP UMR6252, F-14000 Caen, France; \href{https://orcid.org/0000-0001-7217-2707}{orcid.org/0000-0001-7217-2707}

\section{ACKNOWLEDGMENTS}
This research was supported by funding from the French government, managed by the National Research Agency (ANR), under the France Grant No.~2030 program, Reference No.~ANR-23-EXES-0001 and the Normandy Region. This project has received funding from the European Union's Horizon 2020 research and innovation programme under grant agreement No 824096 (RADIATE). 
D.G.P. acknowledges support from the National Science Centre, Poland, grant no. 2022/47/D/ST4/03286, the scholarship awarded from the Polish Ministry of Education and Science to outstanding young scientists (decision no. SMN/18/1668/2022), the generous allocation of computer time at Wroclaw Centre for Networking and Supercomputing (http://www.
wcss.pl), grant No. 1764595765, and COST Action "Carbon molecular nanostructures in space (NanoSpace CA21126)". P. E.-J. acknowledges support from the Swedish Research Council (grant no. 2017-04106). R.F. thanks the Swedish Research Council (grant numbers 2023-03464), the Knut and Alice Wallenberg Foundation (grant numbers 2017.0104 and 2024.0120), Sweden, and the Olle Engkvist Foundation (grant number 2017/96 (180)) for financial support.
We also acknowledge the CIMAP and GANIL technical staff for their support and special thanks to Claire Feierstein for the smooth running of the facility.

\textbf{Supporting Information Available:} Mass spectra for all three projectile ions. Conformational sampling, binding energies, complementary potential energy surfaces with extensively explored transition states for relevant intermediate steps as well as structure and energy resolve H and H$_2$ losses.

\bibliography{Ada_sub}

\end{document}